\documentclass{jfm}
\usepackage{graphicx}
\usepackage{epstopdf, epsfig}
\usepackage{amsmath}
\usepackage{xcolor}

\shorttitle{Solidification during tin drop impact}
\shortauthor{M.V. Gielen et al.}

\title{Solidification of liquid metal drops during impact}

\author{Marise V. Gielen\aff{1,2},
  Ri{\"e}lle de Ruiter\aff{2},
Robin B. J. Koldeweij\aff{1},
Detlef Lohse\aff{1},
Jacco H. Snoeijer\aff{1,3}
 \and Hanneke Gelderblom\aff{1,3}
\corresp{\email{h.gelderblom@tue.nl}}}

\affiliation{\aff{1}Physics of Fluids Group, Max Planck Center Twente for Complex Fluid Dynamics, JM Burgers Center, and MESA+ Center for Nanotechnology, Department of Science and Technology, University of Twente, P.O. Box 217, 7500 AE Enschede, The Netherlands.
\aff{2}ASML The Netherlands B.V., De Run 6501, 5504 DR Veldhoven, The Netherlands
\aff{3}Department of Applied Physics,
Eindhoven University of Technology, Den Dolech 2, 5600 MB, Eindhoven, Netherlands}

\begin{document}

\maketitle

\begin{abstract}
Hot liquid metal drops impacting onto a cold substrate solidify during their subsequent spreading.
Here we experimentally study the influence of solidification on the outcome of an impact event. Liquid tin drops are impacted onto sapphire substrates of varying temperature. The impact is visualised both from the side and from below, which provides a unique view on the solidification process. During spreading an intriguing pattern of radial ligaments rapidly solidifies from the centre of the drop. This pattern determines the late-time morphology of the splat. A quantitative analysis of the drop spreading and ligament formation is supported by scaling arguments. Finally, a phase diagram  for drop bouncing, deposition and splashing as a function of substrate temperature and impact velocity is provided. %, where freezing-induced sticking and splashing are observed.

\end{abstract}

%\begin{keywords}
%\end{keywords}

\section{Introduction}

When a liquid metal drop impacts a substrate with a temperature below the liquid's melting point it rapidly cools down and solidifies. For a liquid metal the timescales involved in cooling and solidification are comparable to the impact time of the drop. As a consequence, the impact dynamics of the drop is significantly altered by solidification \citep{Aziz2000, Pasandideh-Fard2002,Mostaghimi:2002, Mehdizadeh:2004, Chandra2009}.

In many applications drops solidify during the impact event. Solidifying liquid metal drops are used in 3D printing to fabricate precise structures \citep{Vaezi2013,Visser2015, Wang:2016}. In thermal spray coating liquid metal drops are sprayed onto a surface to form a protective solid layer \citep{Pasandideh-Fard2002,Mostaghimi:2002,Fauchais2004}. The solidification of impacting rain drops on e.g. road surfaces \citep{Symons:1997}, aircrafts \citep{Cebeci:2003} or transmission lines \citep{Szilder:2002} can have detrimental effects. In all these applications it is important to understand how liquid solidification alters the impact and spreading dynamics of the drops. %\citep{Kreder2016, Schremb2018}

Drop impact under isothermal conditions already shows a rich and complicated dynamics. Even basic properties such as the maximum spreading, fingering instability and splashing threshold are still subject of debate (see e.g.~\citet{Yarin2006, Josserand2016}). Solidification of the liquid during these processes adds further complexity.
The influence of solidification on drop spreading after deposition onto a substrate (i.e.~with zero impact velocity) has been investigated both experimentally and theoretically \citep{Schiaffino1997, Schiaffino1997b, Tavakoli2014, Ruiter2017}. However, the models developed have not yet been tested for drops with a finite impact velocity. 

When a water drop at room temperature impacts a sufficiently cold substrate it first spreads out before it solidifies, cracks or fragments \citep{Ghabache2016}.
Only when the water is supercooled, solidification takes place on the timescale of the impact event itself \citep{Kong:2015,Schremb2017,Schremb2017c,Schremb2018}. In that case, solidification starts from heterogeneous nucleation at the substrate and is followed by the rapid propagation of a thin ice layer along the surface of the drop and the formation of dendrites \citep{Schremb2018}. 
By contrast, hot liquid metal drops impacting a  substrate with a temperature below the melting point undergo an equilibrium solidification process \citep{Fedorchenko2007}.

The influence of solidification on the impact of (plasma-)sprayed metal drops has been studied by \citet{Pasandideh-Fard2002, Mostaghimi:2002, Dhiman2007, Fedorchenko2007, Chandra2009}. However, in sprays it is difficult to observe isolated impact events of drops with a controlled size and velocity.
Experiments of single drop impacts have focussed on the splat morphology by top-view photography \citep{Pasandideh-Fard1998, Aziz2000, Mehdizadeh:2004, Dhiman2005}, or self-peeling of the splat after a low-Weber number impact \citep{RuiterJ2017}. However, the solidification-limited impact dynamics of liquid metal drops has so far remained unexplored and a systematic, quantitative study is lacking.

Here, we experimentally study tin drop impact onto a sapphire substrate with a controlled temperature below the melting point of tin. 
In \S\ref{tin:exp} a set-up to generate isolated liquid tin drops on demand that allows for simultaneous side- and bottom-view imaging is presented. The outcome of a typical experiment is described in \S\ref{tin:series}.
The maximum drop spreading is determined as a function of substrate temperature and a basic model to explain our findings is presented in \S\ref{tin:max}. The bottom view experiments reveal the growth of ligaments over time starting from early-time solidified corrugations.
As a consequence, the number of ligaments that evolves from the rim of the drop is strongly affected by solidification, as is shown in \S\ref{tin:lig}. In \S\ref{tin:spl} the splashing threshold of the impacting drops as a function of substrate temperature is determined quantitatively. We compare our results to previously reported findings and discuss possible extensions of our basic solidification model in \S \ref{tin:con}.

\section{Experimental method} \label{tin:exp}
\begin{figure*}
\centering\includegraphics[width=0.7\textwidth]{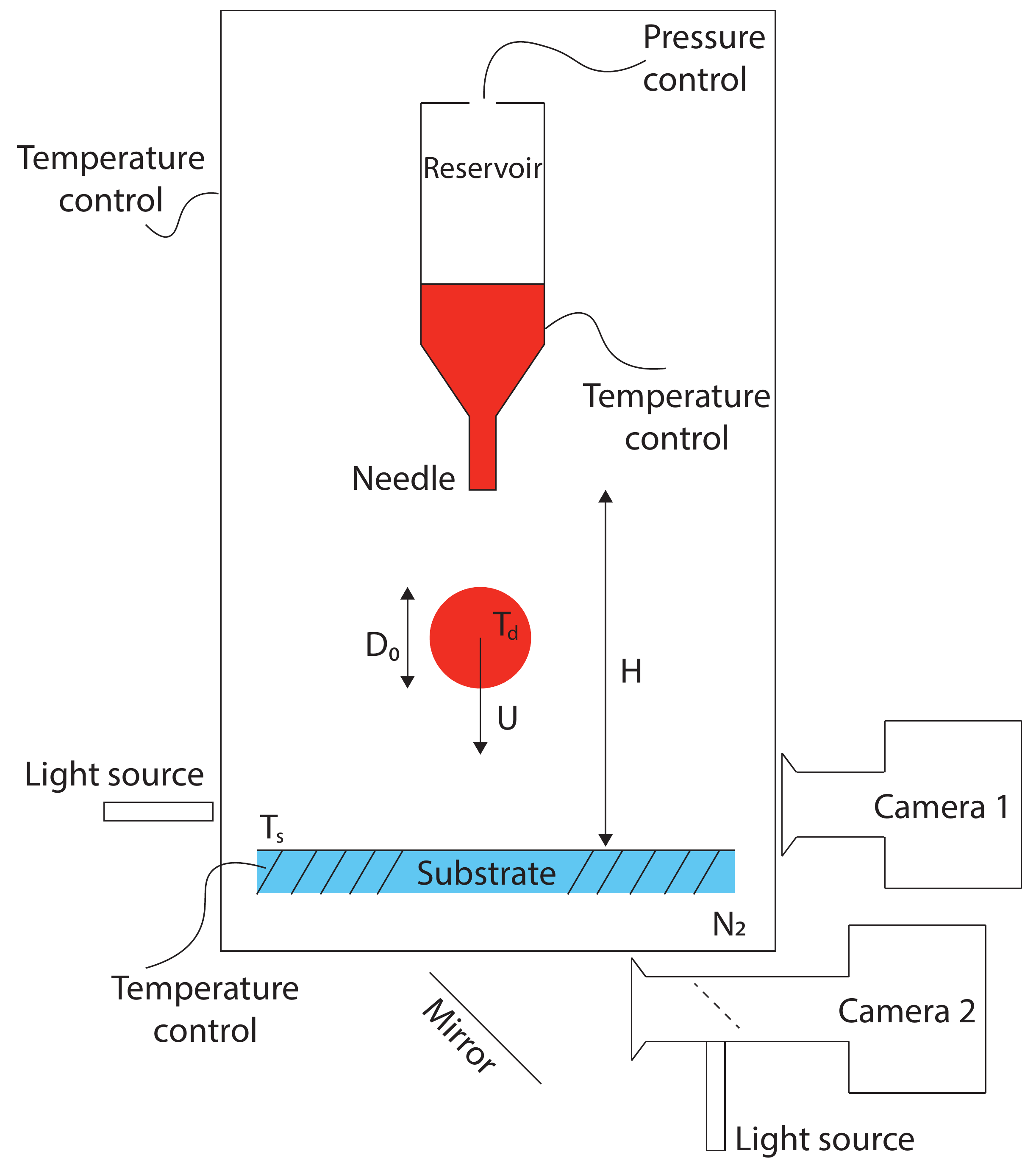}
\caption{Schematic view of the drop impact setup.
Tin is contained in a reservoir that is heated to 250$^\circ$C.
Upon melting the tin fills a needle from which drops of size $D_0=1.7 \pm 0.16$ mm and $2.0\pm 0.43$ mm are pinched.
The reservoir is kept at an underpressure to prevent tin leakage.
To generate a drop a short pressure pulse is given.
The drop falls down under the influence of gravity.
By changing the needle height $H$, the impact velocity is varied from 0.5 ms$^{-1}$ $<U<$ 5.5 ms$^{-1}$.
The temperature of the sapphire substrate $T_s$ is controlled between 40$^\circ$C and 250$^\circ$C.
The substrate and the tin reservoir are placed inside a closed box that is kept at 250$^\circ$C and is filled with nitrogen to prevent cooling and oxidation of the tin drops. The impact event is visualised both from the side and from below.}
\label{fig:tin:exp}
\end{figure*}

Isolated liquid tin drops are impacted onto a sapphire substrate. The choice of sapphire as substrate material is motivated by its transparency and high thermal diffusivity (compared to other transparent materials such as glass), which allows for bottom-view visualisation of the impact event and good control of the substrate temperature.
 In the experiments drop diameter $D_0$, impact velocity $U$ and substrate temperature $T_s$ are varied.
An overview of the setup is shown in Fig.~\ref{fig:tin:exp}.

To create single tin drops on demand we use a method similar to the one described by \citet{Cheng2005, Zhong2014}.
The tin is contained in a reservoir that is heated to $250^\circ$C, which is well above the melting temperature of tin ($T_m = 232^\circ$C). We used 99.9\% pure tin, which has (in liquid phase at $250^\circ$C) a density $\rho = 7.0\cdot10^3$ kgm$^{-3}$, surface tension $\gamma = 0.54$ Nm$^{-1}$, kinematic viscosity $\nu = 2.6\cdot10^{-7}$ m$^2$s$^{-1}$,  thermal conductivity $k=31$ W(mK)$^{-1}$, specific heat $c_p = 2.3\cdot10^2$ J(kgK)$^{-1}$ and latent heat $L = 5.9\cdot10^4$ Jkg$^{-1}$. The bottom of the reservoir is tapered to allow for a smooth connection to the needle (Hamilton, custom made, stainless steel, diameters 0.71 and 1.14 mm).

The pressure inside the reservoir is maintained by controllers that are situated one meter higher, such that they remain at room temperature (working temperature) while the reservoir is heated. The reservoir is kept at an underpressure of $p \approx$ -55 mbar to prevent leakage of liquid tin. To generate a tin drop this underpressure is replaced by a short pressure pulse of 340 mbar (pulse duration between 0.06 s $<\Delta t<$ 0.09 s depending on the height of the tin column in the reservoir).  
The reservoir is mounted on a movable stage of one meter that allows to change the impact height and thereby the impact velocity of the drop between $0.5$ ms$^{-1}$ $<U<5.5$ ms$^{-1}$. The drop size is 1.7 $\pm$ 0.16 mm or 2.0 $\pm$ 0.43 mm, depending on the needle diameter. 

To ensure the liquid tin drop does not oxidise during its fall, the entire setup is contained in a closed box filled with nitrogen gas at a slight overpressure to prevent oxygen leaks.
During heating the setup is flushed with nitrogen while during operation the flow rate is reduced to minimize disturbances to the impacting drop.
The temperature of the box is controlled at $T=250^\circ$C to prevent cooling of the drop during its fall.

At the bottom of the closed box a sapphire substrate (thickness $h = 3$ mm, thermal conductivity k$_{sub}=27$ W(mK)$^{-1}$, specific heat $c_{p,sub}=763$ J(kgK)$^{-1}$ and density $\rho_{sub}=3.98\cdot 10^3$ kgm$^{-3}$) is placed inside a temperature-controlled holder. The centre of the metal holder is open to allow for bottom-view measurements. The substrate temperature is varied between 40$^\circ$C $<T_s<$ 250$^\circ$C. After placement the substrate is left to equilibrate for ten minutes to ensure it is heated uniformly.  

We measure the outcome of the impact using high-speed imaging from the side and from below.
The side-view images (Camera 1 in Fig.~\ref{fig:tin:exp}, Photron SA1.1, 30,000 fps, combined with a long-distance microscope K2 DistaMax by Infinity Photo-Optical Company, resolution 33.8 $\mu$m/pixel) are taken with back-light illumination (Sumita LS-M352A) and the bottom-view images (Camera 2 in Fig.~\ref{fig:tin:exp}, Photron SA-X2, 30,000 fps, combined with a Navitar 12x zoom lens, resolution 18.1 $\mu$m/pixel) with coaxial illumination (Asahi MAX-303).
Both cameras are synchronized to the pressure pulse that generates the drop with a time delay of 0.4 seconds.

\section{Qualitative observations}\label{tin:series}
\begin{figure*}
\centering\includegraphics[width=\textwidth]{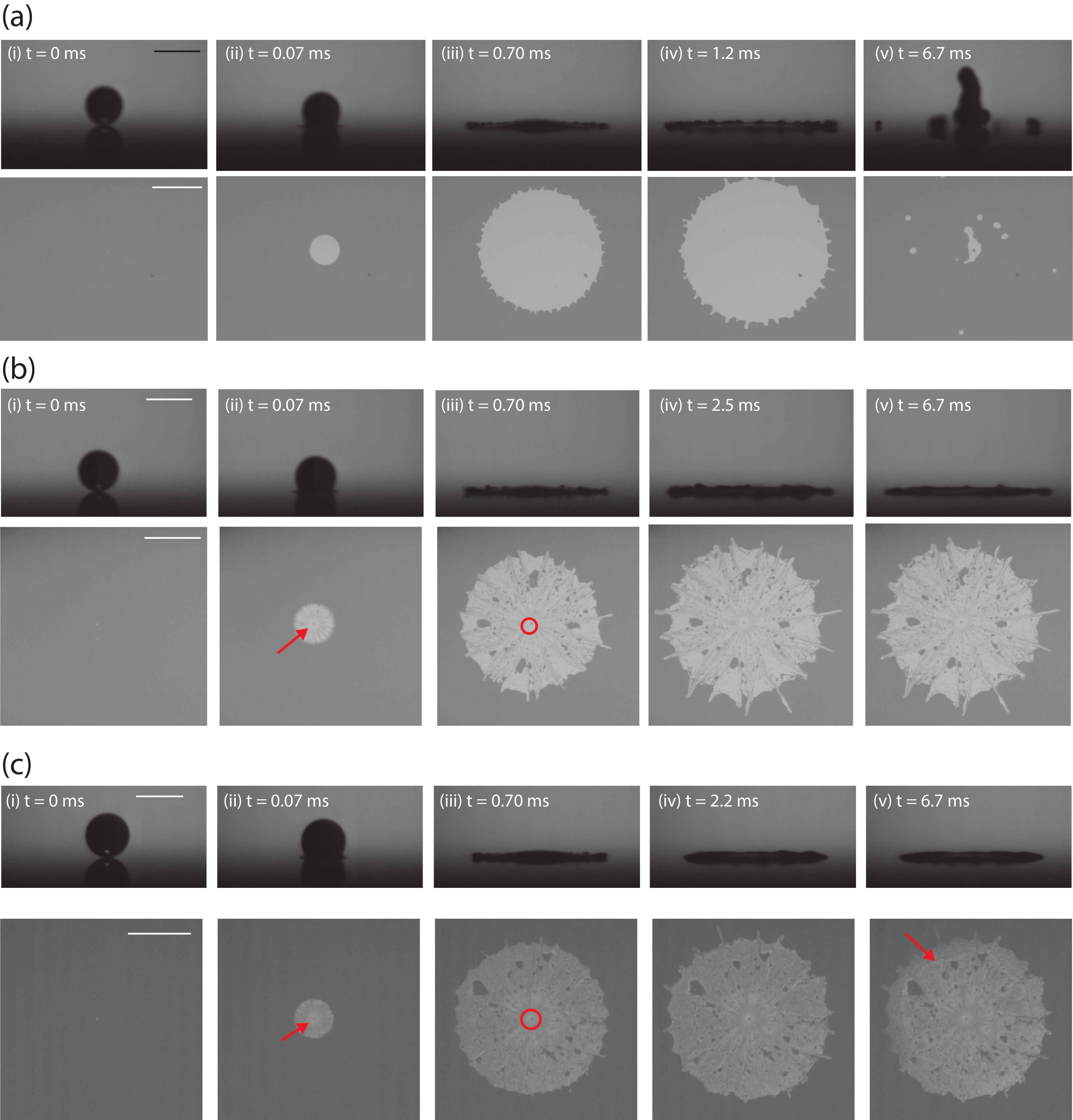}
\caption{Time series of drops  of similar Weber number impacting onto sapphire substrates of different temperatures.
(a) We = 281, $T_s = 249^\circ$C (b) We = 267, $T_s=150^\circ$C, and (c) We = 278, $T_s=39^\circ$C.
The top rows show side-view images, the bottom rows bottom-view images.
The scale bar represents 2 mm.
The drop is shown just before impact (\textit{i}) and spreads over the surface in time (\textit{ii} and \textit{iii}) until it reaches its maximum spreading (\textit{iv}) and retracts or remains solidified at the substrate (\textit{v}). The corresponding movies can be found in the supplementary material.}
\label{fig:tin:series}
\end{figure*}

Figure~\ref{fig:tin:series} shows the time series of three drops impacting at similar Weber number We$=\rho D_0 U^2/\gamma$ onto a sapphire substrate with temperatures of (a) $T_s=$ 249$^\circ$C (isothermal impact), (b) 150$^\circ$C and (c) 39$^\circ$C. 
The top row images show the drop impact event in side view, the bottom row images in bottom view.
Before impact (panel \textit{i} in Fig.~\ref{fig:tin:series}(a-c)) the drop is not visible in bottom view since there is no contact with the substrate. 

During isothermal impact (Fig.~\ref{fig:tin:series}a) the drop spreads smoothly over the substrate at early times (panel a.\textit{ii}).
Over time ligaments become visible (panel a.\textit{iii}), merge and grow until the drop reaches its maximum spreading (panel a.\textit{iv}). As ligaments continue to appear, split and merge their total number is changing over time \citep{Thoroddsen:1998}.  After reaching its maximum spreading the drop retracts and partially bounces off the substrate while small droplets detach (panel a.\textit{v}). This bounce is caused by the repelling nature of the sapphire substrate, which has a contact angle with tin of about 130$^\circ$ (as measured by side-view imaging of a deposited drop).

In Fig.~\ref{fig:tin:series}b the substrate temperature is decreased to $T_s = 150^\circ$C, which is below the melting temperature of tin. While the side view initially looks identical to the isothermal case (cf.~ panels a.(\textit{ii-iv}) and b.(\textit{ii-iv})), the bottom view reveals striking differences: a greyscale pattern appears that remains unchanged as the drop spreads and forms a clear mark of the solidification taking place at the interface with the substrate.  
Directly after impact a grey spot is visible at the impact centre of the drop (red arrow in panel b.\textit{ii}, see also the close-up presented in Fig.~\ref{fig:tin:splat}a) surrounded by a lighter zone and grey radially outward pointing stripes. This central grey spot is caused by the entrapment of an air bubble during first contact between the drop and the substrate \citep{Chandra1991,Dam2004}. In the isothermal impact experiments this air bubble could not be observed long enough in the bottom-view images, presumably due to pinch off and detachment \citep{Lee2012}. The light grey zone surrounding the bubble (red circle in panel b.\textit{iii}, see also Fig.~\ref{fig:tin:splat}b) has no texture. A similar \emph{defect-free} zone $D_d$ has been observed by \citet{RuiterJ2017}. These authors attributed the existence of this zone to a delay in the solidification caused by the vertical flow of hot liquid in combination with the diverging contact line velocity. 

When the drop spreads further (panel b.\textit{iii}) ligaments evolve from its edge. These ligaments immediately solidify on the substrate and elongate as time progresses while their number remains unchanged, in contrast to the isothermal case. The total number of ligaments present at maximum spread is smaller than for the isothermal impact (cf. panels a.\textit{iv} and b.\textit{iv}), as was also observed by \citet{Aziz2000}. In fact, a close inspection shows that the ligaments in the solidified splat can be traced all the way back to the grey stripe-pattern that was already visible at an early stage (panel b.\textit{ii} and Fig.~\ref{fig:tin:splat}c). 
The imprint of the spreading drop also shows dark grey spots.  At these locations the spreading of liquid tin is hindered by imperfections at the substrate, which results in the entrapment of air pockets. 
Due to solidification of the surrounding tin most of these air pockets remain visible at the final splat (panel b.\textit{v}). Solidification reduces the overall spreading velocity of the drop and the maximum spread as compared to isothermal impact. After the maximum expansion is reached the bulk of the drop solidifies (panel b.\textit{v}). A retraction phase is therefore not observed and the bottom-view image no longer changes over time. 

When the substrate temperature  is decreased further ($T_s=39^\circ$C, Fig.~\ref{fig:tin:series}c) we again observe a central air bubble (red arrow in panel c.\textit{ii}),  defect-free zone (red circle in panel c.\textit{iii}), and ligaments that form from a radial stripe pattern (panel c.\textit{iv}).
The number of ligaments is again smaller than for isothermal impact (Fig.~\ref{fig:tin:series}a), but comparable to the experiment shown in Fig.~\ref{fig:tin:series}b. The maximum spreading diameter (panel c.\textit{iv}) of the drop is smaller than in Fig.~\ref{fig:tin:series}b. Moreover, maximum spreading is followed by an inward moving front (red arrow in panel c.\textit{v}), see also supplementary movie). This \emph{self-peeling} of the splat results from a build-up of bending stress due to shrinkage of the tin as it cools down \citep{RuiterJ2017}.  Indeed, the splats of the drops that show this self-peeling are much easier to remove from the substrate after the experiment than those that do not peel off. For impacted water drops, such build-up of bending stress due to solidification has been shown to result in dramatic cracking and fragmentation events \citep{Ghabache2016}.

\subsection{Microscopy of the splat}
\begin{figure*}
\centering\includegraphics[width=\textwidth]{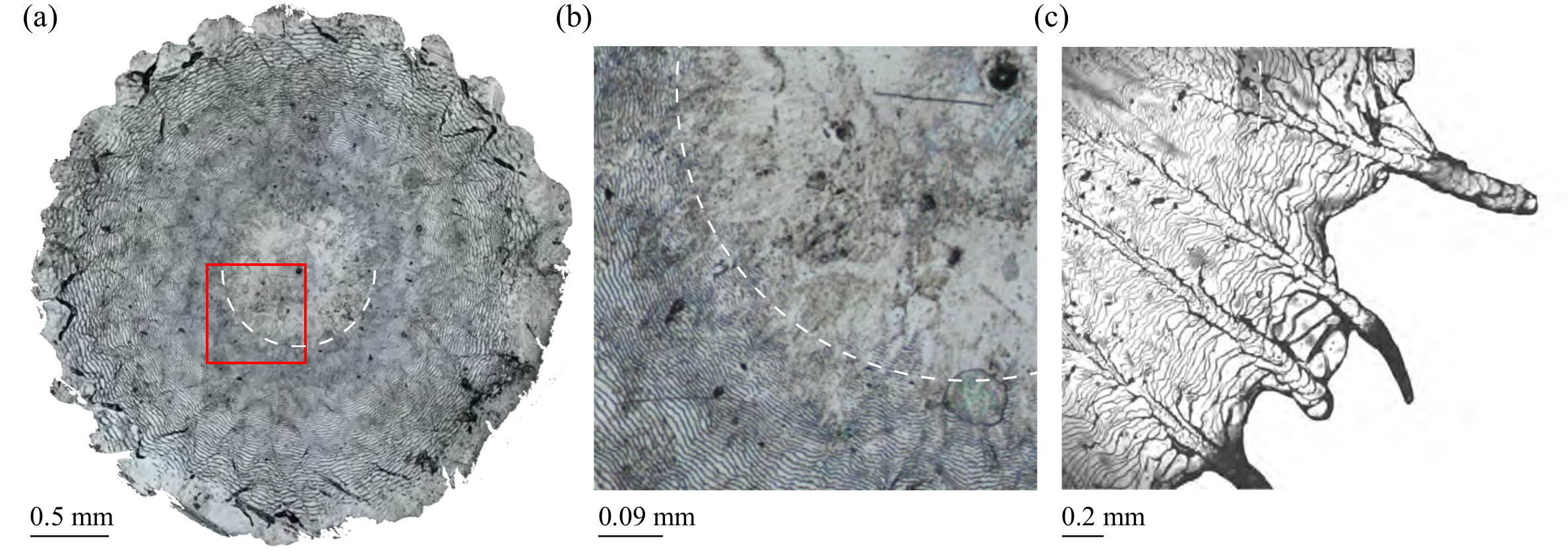}
\caption{Microscope images of the solidified splat of a tin drop after impact onto a sapphire substrate. (a) We~=~135 and $T_s \approx$ 50$^\circ$C. %at 5$\times$ magnification. 
The central bubble (black spot) is surrounded by a lighter defect-free zone. The border of this zone is marked by the dashed white arc. Further outwards, a pattern of circular ridges is observed. (b) Enlargement of the area marked by the red square in panel (a) to illustrate the border of the defect free zone $D_d/D_0\approx 0.6$ (dashed white arc).
(c) We~=~448 and $T_s=150^\circ$C. % 4$\times$ magnification.
Close-up of the splat showing that solidified radial ligaments consist of a sequence of air ridges and can be traced back towards the centre of the splat.}
\label{fig:tin:splat}
\end{figure*}

Splats that self-peel could be removed from the substrate for evaluation under the microscope. An example of a microscope image for an experiment with $\mathrm{We}=135$ is shown in Fig.~\ref{fig:tin:splat}(a,b).
The black spot in the centre of the splat again corresponds to the central air bubble that is trapped in the solidified liquid. The dashed white arc marks the border of the defect-free zone. In the magnified view in panel (b) the defect-free zone and its border can be seen in more detail. Outside this zone a pattern of circular rings is observed, similar to what has been reported by \citet{RuiterJ2017}. This pattern results from the entrapment of air during pinning events at the contact line: when solidification arrests the contact line bulk liquid tin continues to spread over the substrate and forms a new contact. Each time a contact is formed air gets entrapped, which leads to the formation of circular ridges in the splat. 
In this low Weber-number experiment radial stripes and ligaments are not observed. The drop edge however does show undulations with a clear wavelength from which ligaments can evolve once the Weber number is increased.

At larger We, we observe both circular ridge structures, radial stripes and ligaments (see Fig.~\ref{fig:tin:splat}c). One clearly observes that the solidified ligaments consist of a sequence of ridges. This observation suggests that ligaments formed by bulk liquid rapidly flow outward over previously solidified undulations of tin. Thereby, a pattern of radial stripes is formed that can be traced all the way back to the defect-free zone in the centre of the drop.  Hence, the early-time solidification of undulations of the spreading drop determines the fingering pattern observed in the final solidified splat. Note that such structures were not present in the experiments by \citet{RuiterJ2017}, who considered low Weber-number impacts only. 

\section{Drop spreading} \label{tin:max}
\subsection{Experiments}
\begin{figure*}
\centering\includegraphics[width=\textwidth]{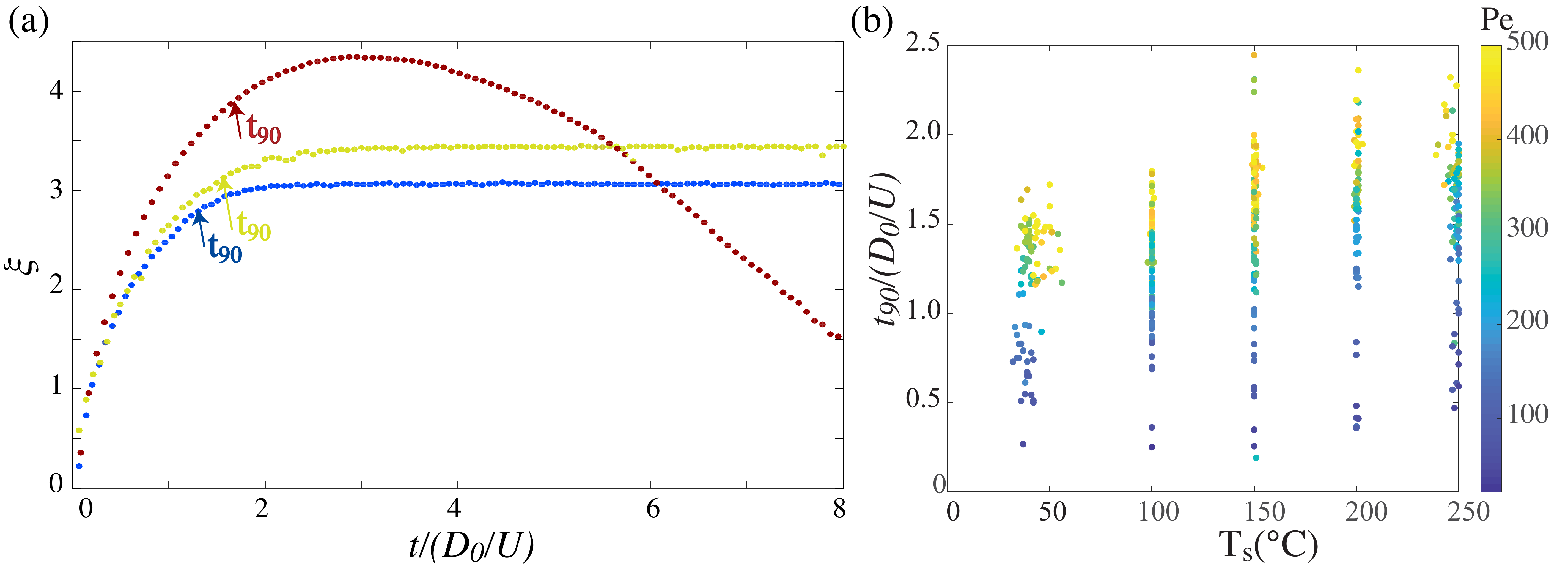}
\caption{(a) Drop spreading factor $\xi=D/D_0$ as function of time $t/(D_0/U)$ for different substrate temperatures at similar impact velocity and size:  $T_s=249^\circ$C, $U=3.9$ ms$^{-1}$, $D_0=1.6$ mm (red dots), $T_s=150^\circ$C, $U=3.5$ ms$^{-1}$, $D_0=1.7$ mm (yellow dots), and $T_s=39^\circ$C, $U=3.5$ ms$^{-1}$, $D_0=1.7$ mm (blue dots). The arrows mark the time $t_{90}$ the spreading diameter reaches 90\% of its maximum value.
%We=310, 267, 278.
(b) Time $t_{90}/(D_0/U)$ as a function of $T_s$ for different values of the Peclet number Pe$=D_0U/\alpha$ (colour bar).}
\label{fig:tin:xitime}
\end{figure*}

We now analyse the spreading dynamics and determine the maximum spreading as function of $U$ and $T_s$.
At each point in time the equivalent diameter $D$ of the area occupied by the spreading drop is determined from the bottom-view images. Figure~\ref{fig:tin:xitime}a shows the typical dimensionless spreading diameter $\xi=D/D_0$ over time $t/(D_0/U)$ for three substrate temperatures at a similar impact velocity: $T_s=249^\circ$C (which is above $T_m$, red dots),  $150^\circ$C (below $T_m$, yellow dots) and $39^\circ$C (blue dots). For each curve the moment in time when the drop diameter reaches $\xi=0.9 \xi_{max}$ is marked by an arrow and labelled as $t_{90}$. As shown in Fig.~\ref{fig:tin:xitime}b $t_{90}/(D_0/U)$ is of order unity, though slightly increases with increasing substrate temperature and increasing Peclet number. 

When the substrate temperature is above the melting point no solidification takes place (red dots in Fig.~\ref{fig:tin:xitime}a). The drop spreads to a maximum diameter $\xi_{max}=D_{max}/D_0$, retracts and bounces off the substrate. We refer to this case as `isothermal spreading'. 

At early times ($t\ll D_0/U$) all drops follow the isothermal spreading curve. As time progresses the spreading of drops on the colder ($150^\circ$C  and $39^\circ$C) substrates slows down and comes to rest before the retraction phase is reached. Hence, $\xi$ remains at its maximum value  $\xi_{max}$.  The colder the substrate, the sooner the spreading deviates from isothermal spreading, the smaller $t_{90}$ and the lower the value of  $\xi_{max}$ achieved.

\begin{figure*}
\centering\includegraphics[width=0.7\textwidth]{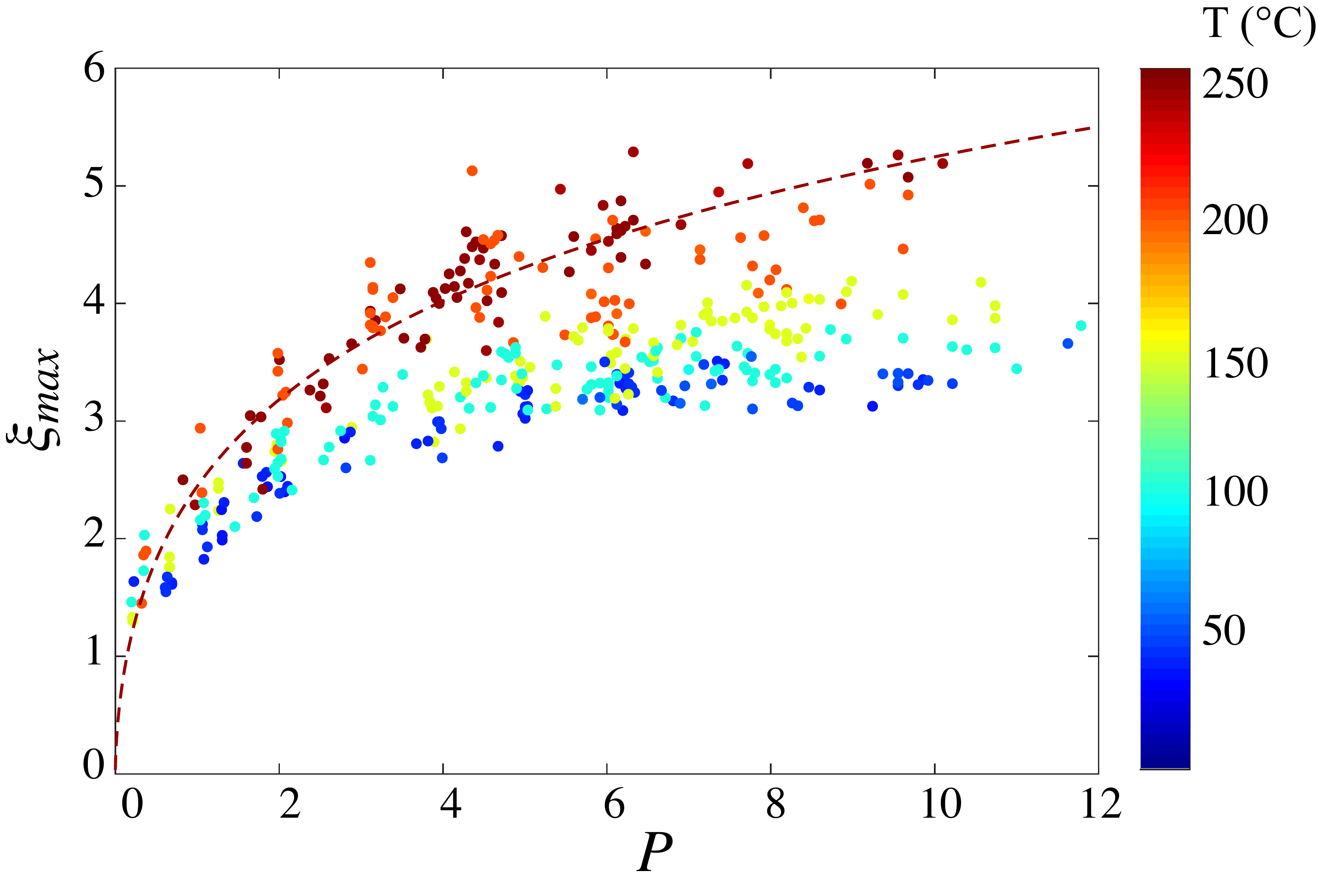}
\caption{Maximum spreading $\xi_{max}$ as function of impact parameter $P=\mathrm{WeRe}^{-2/5}$ for different substrate temperatures: $T_s\approx 250^\circ$C (red dots), $T_s\approx 200^\circ$C (orange dots), $T_s\approx 150^\circ$C (yellow dots), $T_s\approx 100^\circ$C (light blue dots) and $T_s\approx 50^\circ$C (dark blue dots). The dashed red line corresponds to the model (\ref{eq:tin:DmaxRe}) for isothermal spreading by \citet{Laan2014} with a prefactor of 0.9.}
\label{fig:tin:ximax}
\end{figure*}

The maximum spreading diameter reached by the drops is extracted from the curves in Fig.~\ref{fig:tin:xitime}a. For isothermal spreading \citet{Laan2014} proposed the scaling law
\begin{equation}
\xi_{max}\sim\frac{P^{1/2}}{A+P^{1/2}}\textrm{Re}^{1/5}, \label{eq:tin:DmaxRe}
\end{equation}
where $A = 1.24$ is a fitting constant and impact parameter $P=\mathrm{WeRe}^{-2/5}$, with $\mathrm{Re}=UD_0/\nu$ the Reynolds number. In a large part of our experiments $P=\mathcal{O}(1)$, which implies we are at the crossover between the capillary and viscous spreading regime and should use the full Pad\'{e} function to describe our data \citep{Laan2014}. Clearly, given the limited range of data one could also fit $\xi_{max}\sim Re^{1/5}$, but for consistency we retain Eq. (\ref{eq:tin:DmaxRe}).
In Fig.~\ref{fig:tin:ximax} we show the maximum dimensionless spreading $\xi_{max}$ as function of impact parameter $P$ for different substrate temperatures, where we interpret $P$ as the dimensionless impact velocity (control parameter of the experiment) and $\xi_{max}$ as the response of the system.
For isothermal impact our data agrees well with Eq. (\ref{eq:tin:DmaxRe}) when a prefactor of 0.9 is used. This prefactor is somewhat smaller than the prefactor of 1.0 obtained by \citet{Laan2014}. We attribute this small discrepancy to the fact that we determine the spreading diameter from the equivalent area covered by the drop in the bottom-view images, thereby correcting for ligament formation, whereas \citet{Laan2014} used side-view images. 

For substrate temperatures below $T_m$ the maximum spreading is reduced, in particular for larger $P$ and lower $T_s$ (see Fig.~\ref{fig:tin:ximax}). For $T_s\approx 200^\circ$C, which is just below $T_m$, the maximum spreading is close to its isothermal value and only deviates a little for impact parameters larger than $P\approx6$. By contrast, for a drop with $T_s\approx 50^\circ$C and $P=10$ the maximum spreading decreases from $\xi_{max}=5.2$ (isothermal) to $\xi_{max}=3.3$, which is a reduction of 37$\%$. 

\subsection{Model for solidification-limited spreading}
\begin{figure*}
\centering\includegraphics[width=0.7\textwidth]{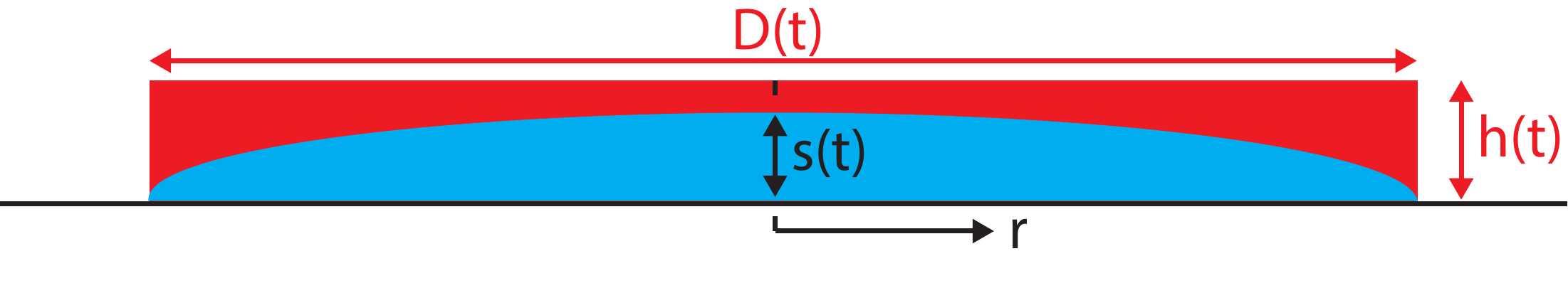}
\caption{Schematic view of the growth of the solidified layer inside the spreading drop.
The drop is assumed to have a pancake shape (in red) with height $h(t)$ and diameter $D(t)$. A solidified layer with thickness $s(t)$ starts growing from the moment the drop contacts the substrate. Arrest occurs when the solidified layer reaches the pancake thickness and bulk liquid can no longer overflow the contact line: $s(t)=h(t)$. Assuming heat transfer in the radial direction $r$ is negligible one obtains a square-root dependence of the solidified layer on $r$ (see text). }
\label{fig:tin:shspread}
\end{figure*}

To explain the behaviour observed in Fig.~\ref{fig:tin:ximax} we derive a scaling relation for drop spreading in a regime where solidification is the limiting factor.
Isothermal spreading dynamics upon impact has successfully been modeled by \citet{Eggers2010}. Inspired by this model \citet{Laan2014} derived the expression (\ref{eq:tin:DmaxRe}) for the maximum spreading. \citet{Eggers2010} assumed that spreading arrests when the thickness of the viscous boundary layer that grows from the substrate equals the height of the drop.
Similarly, here we assume that during solidification-limited spreading the drop arrests when the solidified layer that grows from the cold substrate reaches the drop height, as shown schematically in Fig.~\ref{fig:tin:shspread}. Note that until solidification takes place, the increase in kinematic viscosity of liquid tin due to its cooling is less than $5\%$ \citep{Assael2010} and therefore too small to account for the observed change in the drop spreading.

To find the moment of arrest we first describe the spreading dynamics of the drop. In our experiments arrest occurs at early times $t_a\sim D_0/U$ during the expansion phase (see Fig.~\ref{fig:tin:xitime}b). In this phase the rim still has a negligible mass and the drop shape can well be approximated by a pancake of uniform thickness $h(t)$ \citep{Eggers2010}.
From mass conservation it then follows that the pancake thickness scales as
\begin{equation}
\frac{h}{D_0}\sim \left(\frac{D_0}{D}\right)^{2}. \label{eq:tin:hmass}
\end{equation}
The next step is to find an expression for $D(t)$. 
Explicit scaling relations that cover the entire expansion phase do not exist, as the spreading diameter is the result of a complex interplay between inertial, capillary and viscous forces and eventually also wettability of the substrate \citep{Rioboo2002, Roisman2009, Eggers2010}.  We therefore look for a scaling relation that applies solely to the regime of interest $t\sim D_0/U$. 
At this time viscous effects are confined to a tiny boundary layer of thickness $\delta/D_0\sim \mathrm{Re}^{-1/2}\sim 10^{-2}$ and can be neglected \citep{Riboux2014}. Moreover, as $t_a/t_c\sim \mathrm{We}^{-1/2}\sim 10^{-1}$, with $t_c=\sqrt{\rho D_0^3/\gamma}$ the capillary time scale, we also neglect the influence of surface tension on the spreading dynamics. Since at $t_a$ we have $h\ll D$ we describe the pancake dynamics in the thin-film approximation. The mass and momentum conservation equations then read
\begin{subequations}
\begin{align}
r\frac{\partial h}{\partial t}+\frac{\partial}{\partial r}(ruh)&=0, \label{eq:tin:mass}\\
\frac{\partial u}{\partial t}+u\frac{\partial u}{\partial r} &=0, \label{eq:tin:mom}
\end{align}
\end{subequations}
with $r$ the radial coordinate and $u$ the radial velocity.
Upon integration of (\ref{eq:tin:mass}) and by using (\ref{eq:tin:hmass}), we find $u=r\dot{D}/D$. The momentum balance (\ref{eq:tin:mom}) then reduces to $\ddot{D}=0$, and hence 
\begin{equation}
D\sim Ut. \label{eq:D}
\end{equation}
From (\ref{eq:tin:hmass}) we then find the spreading dynamics of the pancake  for $t\sim D_0/U$ to be given by
\begin{equation}
\frac{h}{D_0}\sim \left(\frac{D_0}{Ut}\right)^{2}. \label{eq:tin:h}
\end{equation}

Second, we model the growth of the solidified layer from the cold substrate. Here we employ a 
one-dimensional heat transfer model similar to \citet{Pasandideh-Fard1998, Aziz2000} based on equilibrium solidification, which is justified in the regime under study \citep{Fedorchenko2007}. The model neglects heat transfer in radial direction and therefore strictly applies to the dynamics of the solidification front $s(t)$ in the centre of the drop, as illustrated in Fig.~\ref{fig:tin:shspread}.
As the drop and its surroundings are kept at approximately $T_m$ the heat transfer from the solidification front occurs solely through the solidified layer. 
In our experiments the Stefan number of the solid 
\begin{equation}
\mathrm{Ste}_s=c_{p,s}\frac{T_m-T_s}{L} \label{eq:ste}
\end{equation}
%, with specific heat $c_{p,s}=243$ J/kgK at temperature of 127$^\circ$C
is typically small (Ste$_s<1$) and therefore the heat transfer is assumed to be quasi steady. The temperature dependence of the thermal properties of solidified tin is neglected and as reference the values at room temperature are used (specific heat $c_{p,s}= 228$ J(kgK)$^{-1}$, density $\rho_s=7310$ kgm$^{-3}$ and thermal conductivity $k_s=67$ W(mK)$^{-1}$). Furthermore, the contact resistance between the sapphire and the tin is neglected \citep{Aziz2000}, which can be justified thanks to the smooth sapphire substrates used in our experiment. The sapphire substrate will, in general, not be isothermal. The interface temperature at the contact is given by the transient convective heat transfer problem between the spreading solidifying tin drop and the conductive sapphire, and lies in between the temperature of the liquid drop and the substrate. To avoid this complexity we use the substrate temperature $T_s$, which is a known experimental input parameter, as a reference temperature in our scaling analysis. The dynamics of the solidification front is then given by \citep[p.185-188]{Bejan1993} 
\begin{equation}
\rho_s L \frac{ds}{dt}=k_s \frac{T_m-T_s}{s},\label{encon}
\end{equation}
with $L$ the latent heat of solidification.
From (\ref{encon}) we find the thickness of the solidified layer to scale as 
\begin{equation}
\frac{s}{D_0}\sim \left(\beta \tilde{t}\frac{\mathrm{Ste}_s}{\mathrm{Pe}}\right)^{1/2}, \label{eq:tin:sstar}
\end{equation}
with $\tilde{t}=t/(D_0/U)$ and $\beta=\alpha_s/\alpha=2.1$ the ratio of  solid and liquid thermal diffusivities and Pe$=UD_0/\alpha$ the Peclet number. Solidification at each radial distance $r$ from the impact centre starts as soon as the liquid contacts the substrate. Neglecting heat transfer in radial direction and using (\ref{eq:D}) and (\ref{eq:tin:sstar}) we find the radial profile of the solidified layer to scale as $s(r,t)\sim\left[D_0(D(t)/2-r)\right]^{1/2}$, as illustrated in Fig.~\ref{fig:tin:shspread}.

We assume that the spreading arrests as soon as there is no bulk liquid left to overflow the solidified liquid at the contact line of the drop. In terms of our scaling model this condition translates to $h(t_{a})=s(t_{a})$. Combining (\ref{eq:tin:h}) and (\ref{eq:tin:sstar}) one finds for the time of arrest
$t_{a}/(D_0/U)\sim\left( \textrm{Pe}/\beta\textrm{Ste}_s\right)^{1/5}$. This scaling of $t_a$ motivated us to  show the dependence of $t_{90}$ on the impact parameters in Fig.~\ref{fig:tin:xitime}b using Pe.
Combining the scaling for $t_a$ with $D_{max}\sim Ut_{a}$ one obtains
\begin{equation}
\xi_{max}\sim\left(\frac{\textrm{Pe}}{\beta\textrm{Ste}_s}\right)^{1/5}. \label{eq:tin:DmaxPe}
\end{equation}
In the regime where $P\gg 1$ and hence the isothermal data follow the scaling $\xi_{max}\sim \mathrm{Re}^{1/5}$, relation (\ref{eq:tin:DmaxPe}) can be obtained straightforwardly: replacement of the kinematic viscosity by the diffusion coefficient of the solidification front propagation $\alpha_s \mathrm{Ste}_s$ directly results in (\ref{eq:tin:DmaxPe}).
 
The scaling relation (\ref{eq:tin:DmaxPe}) is valid as long as the spreading is solidification-limited, which requires that the solidified layer grows faster than the viscous boundary layer and hence $\alpha_s \mathrm{Ste}_s>\nu$.
In addition, surface tension forces should have a negligible influence on the maximum spreading, which requires $t_a<t_c$   or $\mathrm{Pe}/\beta\mathrm{Ste}_s<\mathrm{We}^{5/2}$.
%, which satisfied for our experiments with $\mathrm{Pe}/\beta\mathrm{Ste}_s\mathrm{We}^{5/2}\sim 10^{-1}$. 
When these conditions are not met and longer spreading times are considered scaling (\ref{eq:tin:h}) is no longer valid and the full spreading dynamics of the drop needs to be taken into account.

\begin{figure*}
\centering\includegraphics[width=\textwidth]{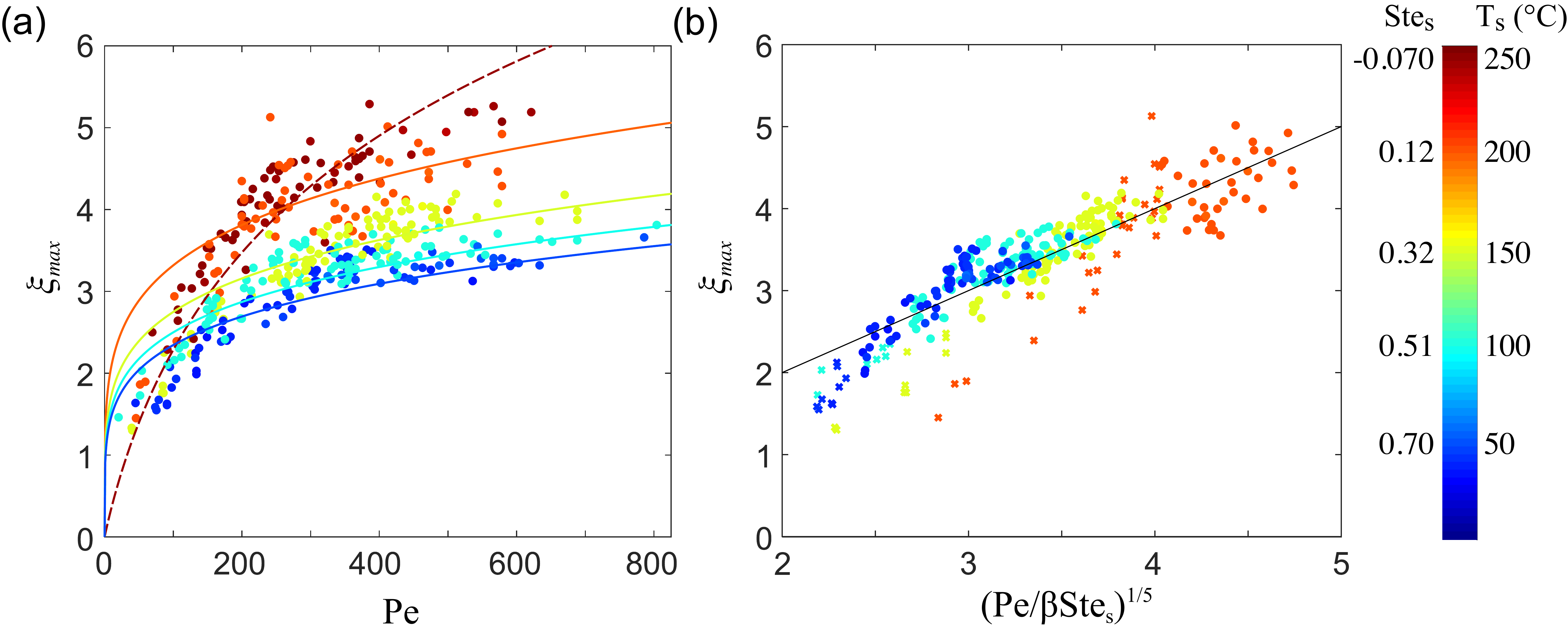}
\caption{(a) Maximum spreading $\xi_{max}$ as function of Pe for different Ste$_s$: Ste$_s\approx -0.070$ (corresponding to $T_s\approx 250^\circ$C, red dots),  Ste$_s\approx 0.12$ ($T_s\approx 200^\circ$C, orange dots),  Ste$_s\approx 0.32$ ($T_s\approx 150^\circ$C, yellow dots),  Ste$_s\approx 0.51$ ($T_s\approx 100^\circ$C, light blue dots) and  Ste$_s\approx 0.70$ ($T_s\approx 50^\circ$C, dark blue dots). The solid lines correspond to scaling (\ref{eq:tin:DmaxPe}) for solidification-limited spreading with a prefactor unity.
The dashed line is the isothermal spreading (\ref{eq:tin:DmaxRe}) rewritten in terms of Pe with prefactor 0.9.
(b) When rescaled in terms of $\left(\mathrm{Pe}/\beta\mathrm{Ste}_s\right)^{1/5}$ for $\mathrm{Ste}_s>0$ all data collapse onto a single curve given by (\ref{eq:tin:DmaxPe}) with a prefactor unity (solid line). The data points marked by crosses are not solidification limited and therefore do not follow the master curve (see text).}
\label{fig:tin:Dmax}
\end{figure*}
In Fig.~\ref{fig:tin:Dmax} scaling model (\ref{eq:tin:DmaxPe}) is compared to the experimental data. 
Figure~\ref{fig:tin:Dmax}a shows the same data as in Fig.~\ref{fig:tin:ximax} but now plotted in terms of Pe for different values of Ste$_s$. For Ste$_s>0$ scaling relation (\ref{eq:tin:DmaxPe}) is shown with a prefactor of unity (solid lines). In addition, we also show the isothermal scaling (\ref{eq:tin:DmaxRe}) with prefactor 0.9 (red dashed line). Note that by rescaling (\ref{eq:tin:DmaxRe}) in terms of Pe we introduce an explicit $D_0$-dependence, which results in a somewhat poorer collapse of the isothermal experimental data. Scaling (\ref{eq:tin:DmaxPe}) is in good agreement with the experimental data. As expected for smaller Pe a deviation is observed, in particular at smaller Ste$_s$ (higher substate temperatures). For these experiments (\ref{eq:tin:DmaxRe}) predicts a smaller spreading than (\ref{eq:tin:DmaxPe}) and hence solidification is not the rate-limiting factor, but viscous and/or surface tension effects dominate such that the data actually follows (\ref{eq:tin:DmaxRe}).

To further validate (\ref{eq:tin:DmaxPe}) we plot all data with Ste$_s>0$ rescaled in terms of $\left(\mathrm{Pe}/\beta\mathrm{Ste}_s\right)^{1/5}$ as shown in Fig.~\ref{fig:tin:Dmax}b. The experiments where solidification is not rate-limiting are marked by crosses. For each Ste$_s$ the data corresponding to solidification-limited spreading now collapse onto a single master curve given by (\ref{eq:tin:DmaxPe}) with a prefactor of unity. 

\section{Number of ligaments} \label{tin:lig}
\subsection{Experiments}
\begin{figure*}
\centering\includegraphics[width=0.7\textwidth]{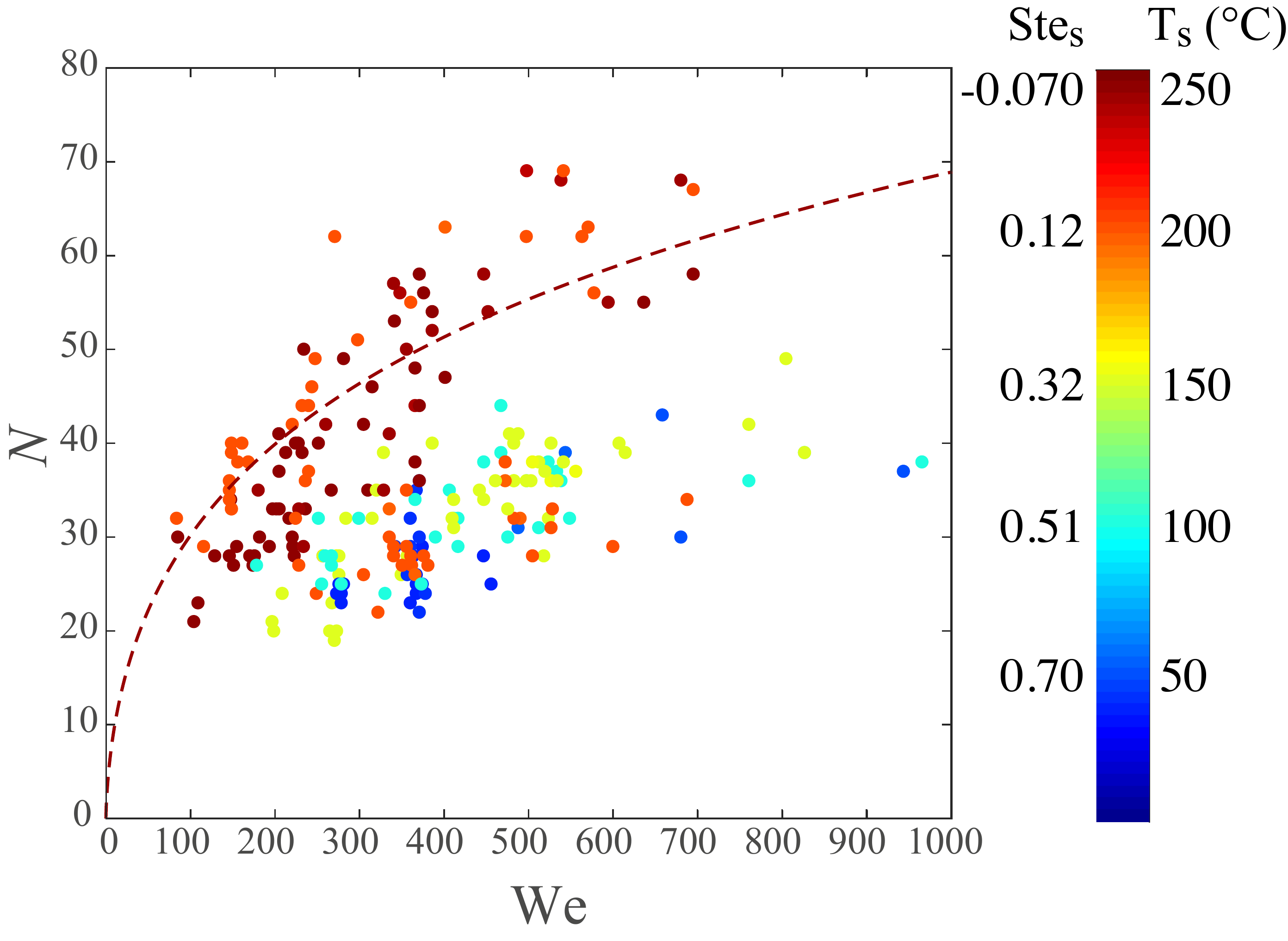}
\caption{Number of ligaments $N$ as function of We for different Ste$_s$: Ste$_s\approx -0.070$ (corresponding to $T_s\approx 250^\circ$C, red dots),  Ste$_s\approx 0.12$ ($T_s\approx 200^\circ$C, orange dots),  Ste$_s\approx 0.32$ ($T_s\approx 150^\circ$C, yellow dots),  Ste$_s\approx 0.51$ ($T_s\approx 100^\circ$C, light blue dots) and  Ste$_s\approx 0.70$ ($T_s\approx 50^\circ$C, dark blue dots).
The red dashed line shows the number of ligaments for isothermal impact as predicted by (\ref{eq:niso}) with a prefactor of 4.3.
For We~$\lesssim$~150 no ligaments are observed.}
\label{fig:tin:N}
\end{figure*}

In Fig.~\ref{fig:tin:series} we observed that solidification has a strong influence on both the development of the ligaments and their final number.
To quantify these observations we manually count the number of ligaments $N$ as a function of the Weber number for different substrate temperatures, as shown in Fig.~\ref{fig:tin:N}. For isothermal impact ligaments appear, split and merge as time progresses (see also \S\ref{tin:series}). To find $N$ we therefore count the number of ligaments at the moment the drop reaches its maximum expansion, before retraction of the rim causes merger. This ambiguity in the definition of $N$ does not appear when the drop solidifies during impact and the number of ligaments is constant.

As is clear from Fig.~\ref{fig:tin:N}, no ligaments are observed for We~$\lesssim$~150. For such small We the impact is not violent enough and therefore the rim retracts before ligaments become apparent. 
For larger Weber numbers $N$ increases with increasing We and is spread in two clouds: isothermal impacts ($T_s>T_m$, red data points), and solidification-limited impacts ($T_s<T_m$, yellow and blue data points). The orange data points correspond to experiments on substrates close to $T_m$ and appear in both clouds. 

Even though each cloud has a considerable spread in $N$, we systematically observe that solidification leads to fewer ligaments. For example, $N$ (almost) halves when we compare the impact onto the coldest substrate ($T_s=50^\circ C$) to isothermal impact ($T_s=250^\circ$C).

\subsection{Qualitative explanation}
To predict the number of ligaments formed during isothermal impact several models based on a Rayleigh-Taylor instability of the decelerating rim have been proposed, see e.g.~\citet{Allen:1975, Marmanis:1996,Kim:2000, Aziz2000, Villermaux2011}. In these models complicated effects such as the splitting and merging of fingers \citep{Thoroddsen:1998} and the influence of ambient pressure \citep{Xu2005, Xu2007} or surface roughness \citep{Range:1998} have been neglected. Despite these shortcomings we now derive a modified version of the model by \citet{Aziz2000} as a guide-to-the-eye to describe the isothermal data shown in Fig.~\ref{fig:tin:N}. 

A Rayleigh-Taylor instability of the rim gives rise to a preferred wavenumber that scales as $k \sim\sqrt{\rho (-\ddot{D)}/\gamma}$, where $\ddot{D}$ is the acceleration of the rim. The number of ligaments is then given by $N\sim kD_\ell$, where $D_\ell$ is the spreading diameter at the moment of ligament expulsion. For isothermal impact ligaments are expelled from the rim close to the maximum drop expansion, such that $D_\ell\sim D_{max}$ \citep{Aziz2000}. At that moment, we estimate the rim deceleration as $\ddot{D}\sim D_{max}/t_c^2$ \citep{Villermaux2011} and hence $k\sim \sqrt{D_{max}/D_0^3}$ . 
Using the maximum expansion given by (\ref{eq:tin:DmaxRe}) we find
\begin{equation}
N\sim \left(\frac{P^{1/2}}{A+P^{1/2}}\textrm{Re}^{1/5}\right)^{3/2}\label{eq:niso}
\end{equation}
 In the limit of  small $P$ (large Re) one recovers that $N\sim \mathrm{We}^{3/4}$  as was derived by \citet{Villermaux2011} for drops impacting on a pillar. Scaling (\ref{eq:niso}) differs somewhat from the one proposed by \citet{Aziz2000} who used $U^2/D_0$ as estimate for the rim deceleration as well as a different expression for $D_{max}$. In Fig.~\ref{fig:tin:N} we show that (\ref{eq:niso}) with a prefactor of 4.3 describes the isothermal experimental data.

For impacts on cold substrates $N$ falls below the isothermal curve (\ref{eq:niso}). This decrease in $N$ was already observed from top-view photography of tin splats by \citet{Aziz2000}, who attributed it to the smaller maximum spreading $D_{max}$ for solidification-limited impacts. Our bottom-view images however allow us to follow the growth of the ligaments over time and reveal a remarkable feature: for the solidified drops the number of ligaments is set by the early-time freezing of undulations and remains unchanged during the subsequent drop spreading, thereby forming a pattern of radially outward pointing stripes (recall Figures \ref{fig:tin:series} and \ref{fig:tin:splat}c).
By contrast, for isothermal impact these early-time undulations remain invisible and ligaments evolving from the rim split at later times \citep{Thoroddsen:1998}, thereby increasing $N$ (compare Fig.~\ref{fig:tin:series}(a.\textit{iv}) with Fig.~\ref{fig:tin:series}(c.\textit{ii})). These two mechanisms qualitatively explain the two clouds of data for isothermal and solidification-limited spreading in Fig.~\ref{fig:tin:N}. For temperatures close to the melting temperature of tin (orange data in Fig.~\ref{fig:tin:N}) both mechanisms are at play: in some drops solidification is fast enough to prevent splitting of the fingers, while in other drops new ligaments emerge at later times.

\section{Splashing threshold} \label{tin:spl}
\subsection{Experiments}
\begin{figure*}
\centering\includegraphics[width=\textwidth]{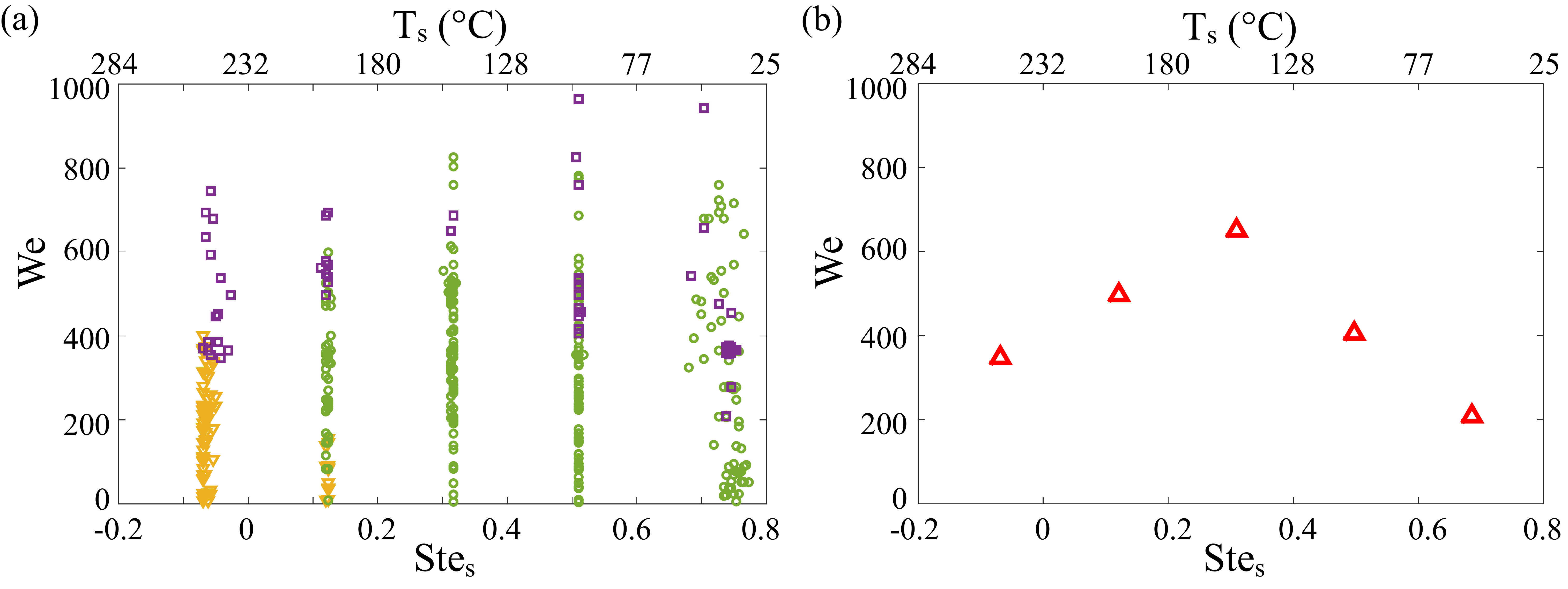}
\caption{(a) Phase diagram showing the outcome of drop impact as function of We and Ste$_s$.
The three impact behaviors observed are bouncing (orange triangles), deposition (green circles) and splashing (purple squares).
(b) Plot of the first splashing event observed as function of We and Ste as extracted from the data in Fig.~\ref{fig:tin:phase}(a).}
\label{fig:tin:phase}
\end{figure*}

An impact event may have several outcomes, depending on the impact conditions. We observed deposition of the drop, which means that the drop remains intact and adheres to the substrate after impact. Splashing occurs when secondary droplets are generated. Bouncing takes place when (part of) the drop lifts off from the substrate.
In Fig.~\ref{fig:tin:phase}a we quantify these impact phenomena as a function of the Stefan and Weber numbers, i.e.~the dimensionless substrate temperature and impact velocity.

For the isothermal impact events ($\mathrm{Ste}_s\leq 0$, $T\geq T_m$) and We~$<350$ the drop bounces. For We~$>$~375 drops always splash during isothermal impact. For intermediate Weber numbers a transition regime exists where both bouncing and splashing can occur. Drop deposition is not observed in the isothermal regime. When the substrate temperature is decreased below the melting temperature (Ste$_s$=0.12, $T_s=200^\circ$C) only a few bounces occur at low We, while most of the drops deposit and stick to the substrate. For higher Weber numbers we again observe a transition to splashing. On even colder substrates (Ste$_s>0.12$, $T_s<200^\circ$C) the bouncing regime completely disappears and we only observe \emph{freezing-induced sticking} of the drop.
The Weber number at which the transition from bouncing or deposition to splashing takes place varies with the Stefan number. Importantly, for solidification-limited impacts there is no longer a sharp boundary between bouncing/deposition and splashing, but both can be observed even at the highest Weber numbers accessible by our setup.

The splashing threshold is analysed in more detail in Fig.~\ref{fig:tin:phase}b, where the critical Weber numbers corresponding to the first splashing event (i.e.~the lowest We where splashing is observed) are plotted as a function of Ste$_s$. The splashing threshold shows a non-monotonous trend: it first increases up to Ste$_s$~$\approx$~0.3 and then decreases again, to a value that lies below the isothermal splashing threshold.

\subsection{Qualitative explanation}
%To explain the behaviour shown in Fig.~\ref{fig:tin:phase}b, we note that the generation of secondary droplets (i.e.~splashing) may occur during several stages of the drop spreading and has several causes: the early-time destabilisation of the ejecta sheet (which can result in a crown splash if the ejecta lifts upwards), the breakup of the advancing contact line due to its interaction with surface roughness (the prompt splash) and the late-time breakup of ligaments formed at the rim \citep{Rioboo2001,Yarin2006, Xu2007b}. In our experiments splashing is also observed at different times. However, in most experiments it is difficult to distinguish between the different types of splashing. In the data for the splashing threshold presented in Fig.~\ref{fig:tin:phase} we therefore do not make this distinction. 

To explain the observations in Fig.~\ref{fig:tin:phase}b, we hypothesize that the influence of solidification on the splashing threshold is twofold. First, solidification limits the spreading of the drop and the growth of ligaments as was observed in \S \ref{tin:max} and \S\ref{tin:lig}. Consequently, one expects splashing due to ligament breakup to be suppressed, as was already suggested by \citet{Aziz2000}. This effect becomes stronger with increasing Ste$_s$ (lower substrate temperatures), which leads to the observed increase in the critical Weber number for the first splash.

Second, rapid solidification of the ejecta sheet forms local irregularities on the substrate, thereby increasing the roughness as discussed in \S\ref{tin:lig}. These irregularities interact with the moving contact line and may thereby provoke a \emph{freezing-induced splash} \citep{Dhiman2005, Dhiman2007, Chandra2009}. This effect becomes more severe with increasing Ste$_s$ such that at some point the splashing threshold decreases below the one for isothermal impact.  As the freezing-induced splash depends on nucleation events at the substrate it has a stochastic nature and is not observed in each experiment.

These two competing effects, the splash suppression by solidification and the freezing-induced splash, can qualitatively explain the non-monotonous trend in the splashing threshold observed in Fig.~\ref{fig:tin:phase}b. 

\section{Discussion and conclusion} \label{tin:con}
The influence of solidification on the impact dynamics of liquid tin drops was investigated. 
Solidification was shown to strongly alter drop spreading, destabilization through the formation of ligaments and splashing. The use of transparent sapphire substrates allowed for bottom-view imaging and provided a unique view on the solidification and impact dynamics. We now discuss our findings in light of previously reported results and reflect on possible extensions for our work.

Spreading of the liquid tin drops on cold substrates is limited by solidification rather than by surface tension or viscous effects. The spreading arrests early in the expansion phase, which results in a smaller maximum spreading. In previous studies the maximum drop spreading was modelled based on energy arguments that account for a loss in kinetic energy due to solidification \citep{Pasandideh-Fard1998, Dhiman2005, Dhiman2007}. In these models arrest was assumed to occur at fixed time, independent of the substrate temperature. Our measurements of drop spreading over time are not consistent with this view, as we found the arrest time to decrease with decreasing substrate temperature. Instead, we derived a model for the solidification-limited spreading dynamics assuming that arrest occurs once the growing solidified layer reaches the height of the spreading drop. From the predicted time of arrest we then found the maximum spreading diameter. In this model we used a basic description of the drop dynamics, neglecting the growth of a viscous boundary layer and surface tension forces acting on the rim of the drop. This simple model already gave a good prediction for the maximum drop spreading. To accurately describe the solidification-limited spreading dynamics a more sophisticated model is needed, for example an extension of the spreading model derived by \citet{Eggers2010} to account for the growth of a solidified layer. Such a model would provide a complete description of drop spreading in presence of solidification and could account for the gradual deviation from isothermal spreading we observed in our data.

In a previous study with hexadecane drop spreading during deposition (i.e. with zero impact velocity) by \citet{Ruiter2017} kinetic undercooling effects were used to obtain a criterion for contact line arrest. In that work, the entire drop was assumed to arrest abruptly once the liquid at the contact line reaches a critical temperature that is determined by kinetic undercooling. Our microscope images of the solidified splat revealed a sequence of circular ridges formed at the interface with the substrate. The moment the first ridge is observed and hence the diameter of the defect-free zone are consistent with the kinetic undercooling scenario, as explained in Appendix~\ref{app:tin:clvel}. 
This observation, together with the gradual arrest of the drop spreading suggests that a sequence of contact line arrests occurs, each time overflown by fresh liquid from the bulk, until the bulk of the drop is solidified and the drop reaches its maximum spreading diameter.

The early-time solidification of the contact line was shown to control the late-time fingering instability: our bottom-view images revealed a pattern of radial stripes that could be traced back all the way to the defect-free zone. These stripes in turn consisted of a sequence of solidified ridges starting from the very early stages of impact.
This finding is consistent with the observation by \citet{Thoroddsen:1998} that the fundamental number of disturbances forms right after the first contact of the drop with the substrate and remains unchanged during the expansion phase. For isothermal drops these fundamental ligaments undergo splitting and merging at later times. By contrast, we have shown that for drops impacting cold substrates the initial undulations solidify and perturb the flow of the remaining liquid, thereby controlling the number of ligaments during the entire expansion phase. The pattern of solidified undulations will depend on the interaction between the substrate, the moving and destabilising contact line and local nucleation events. Therefore, a quantitative prediction of the fingering instability in presence of solidification remains an open issue.

The solidification of early-time undulations of the contact line are also a potential cause of the freezing-induced splashing we observed. The solidified pattern locally increases the surface roughness and can thereby trigger instabilities in the spreading drop.  As a result, splashing is observed at Weber numbers below the isothermal splashing threshold.  
By contrast, solidification can also cause deposition to occur at Weber numbers above the isothermal splashing threshold by suppressing the ligament breakup. To further quantify these effects an experimental setup that can achieve higher Weber numbers is required, possibly through a different drop generation method. Moreover, a clarification of the splash-suppressing and -enhancing effects of solidification requires detailed experimental data to distinguish between splashing caused by contact line destabilisation and by ligament breakup.

The quantitative prediction of the occurrence of the freezing-induced splash for solidifying drops remains challenging. On the theoretical side, one would need to account for the interaction of the ejecta sheet dynamics, the moving contact line, the growing solidification front and the underlying substrate, for example by extending the model by \citet{Riboux2014}. On the experimental side, one would need detailed data of the early stages of impact to disentangle the order in which nucleation, contact line destabilisation, arrest and sheet ejection take place.

Despite these challenges, the solidification of a liquid drop during an impact event  offers a unique perspective on the impact dynamics.
For example, the detailed inspection of the splat formed after impact can be used as a diagnostic tool to identify the impact conditions, such as impact speed and angle \citep{Laan2014, Gielenthesis}. Moreover, the visualization of the impact of solidifying drops forms a powerful method to learn about isothermal impact as one literally freezes the dynamics. The freezing pattern that develops at the solid-liquid interface reveals the different stages of impact, from bubble entrapment to sheet ejection, the growth of corrugations and the development of ligaments. 

\begin{acknowledgements}
We thank Michel Riepen for fruitful discussions.
This work is part of an Industrial Partnership Programme
of the Netherlands Organization for Scientific Research (NWO). This
research programme is co-financed by ASML.
\end{acknowledgements}

\appendix

\section{}\label{app:tin:clvel}
%\section{Contact line arrest criterion for impacting drops} \label{app:tin:clvel}
Here, we derive a criterion for the first contact-line arrest that occurs in solidifying spreading drops after impact. This first arrest determines the size of the defect free zone, as observed in (e.g.) Fig.~\ref{fig:tin:splat}.  To this end, we modify the arrest criterion for drops spreading after deposition described by \citet{Ruiter2017} to account for the impact dynamics. 

According to \citet{Ruiter2017} spreading arrests abruptly when the solidification front catches up with the moving contact line. For rapidly moving contact lines the velocity of the solidification front is limited by the crystallisation rate, which in turn depends on the amount of kinetic undercooling of the liquid \citep{Amini2006}. 
For a small amount of kinetic undercooling the solidification front velocity $U_{f}$ is given by \citep{Fedorchenko2007}
\begin{equation}
U_{f}=\kappa \left(T_m-T_{f}\right), \label{eq:tinapp:tfr}
\end{equation}
with $T_{f}$ the temperature of the solidification front and $\kappa$ the kinetic coefficient. This coefficient is a property of the liquid that is available in the literature for some metals (see e.g.~\citet{Rodway1991,Fedorchenko2007}) but unfortunately has so far not been determined for tin.

The velocity of the contact line follows from the isothermal spreading dynamics of the drop. 
In our experiments the drop diameter at the first arrest $D_d$ is small ($D_d\ll D_{max}$, see Fig.~\ref{fig:tin:splat}a). In this early-time kinematic regime ($t<0.1$) surface tension and viscous effects do not yet affect the spreading dynamics and the drop diameter is given by \citep{Rioboo2002}
\begin{equation}
\xi=c \tilde{t}^{1/2}, \label{eq:tinapp:xi}
\end{equation}
where $\tilde{t}=t/(D_0/U)$ and the proportionality constant $c=\sqrt{6}$ \citep{Riboux2014}.
The contact line velocity then reads
\begin{equation}
U_{cl} =\frac{1}{2}U\frac{d\xi}{d\tilde{t}}= \frac{1}{4}cU\tilde{t}^{-1/2} =\frac{1}{4} c^2U\xi^{-1}. \label{eq:tinapp:ucl}
\end{equation}

In line with \citet{Ruiter2017} we assume that the contact line arrests once the velocity of the contact line equals the velocity of the solidification front; $U_{cl}=U_{f}$.
Hence, arrest occurs once the temperature of the contact line is equal to the temperature of the solidification front, i.e.~$T_{cl}=T_{f}$.
For small contact angles temperature gradients inside the drop are negligible such that the temperature at the contact line equals the substrate temperature $T_{cl}=T_s$ \citep{Ruiter2017}.
The arrest criterion is then given by
\begin{equation}
U_{cl}=\kappa (T_m-T_s). \label{eq:tinapp:crit}
\end{equation}

Combining (\ref{eq:tinapp:ucl}) and (\ref{eq:tinapp:crit}) we obtain an expression for the defect distance
\begin{equation}
\xi_d=\frac{D_d}{D_0} = \frac{1}{4}c^2\frac{U}{\kappa(T_m-T_s)}.\label{xid}
\end{equation}

For the experiment shown in Fig.~\ref{fig:tin:splat}(a,b) we find $\xi_d\approx 0.6$ (as determined from high-resolution images). Relation (\ref{xid}) then gives $\kappa=0.034$ m(sK)$^{-1}$, which is of the right order of magnitude for metals \citep{Rodway1991,Fedorchenko2007, Monk2010}. Our analysis therefore provides evidence that the first contact line arrest, and hence the size of the defect-free zone is determined by non-equilibrium crystallization.

\bibliographystyle{jfm}

\bibliography{bib}

\end{document}